\documentclass{basi}
\usepackage[T1]{fontenc}
\usepackage[british]{babel}
\usepackage[varg]{txfonts}
\begin{document}

\title[Environmental dependence of galaxy age in DR10]{Environmental dependence of galaxy age in the Main galaxy sample of SDSS DR10}

\author[X. F.\ Deng ]{Xin-Fa Deng\thanks{Current address: School of Science, Nanchang University, Jiangxi, China, 330031, email: \texttt{xinfadeng@163.com}} \\
    School of Science, Nanchang University, Jiangxi, China, 330031}

\pubyear{2014}
\volume{42}
\pagerange{\pageref{firstpage}--\pageref{lastpage}}

\date{Received 2014 May 05; accepted 2014 August 29}

\maketitle

\label{firstpage}

\begin{abstract}
Using two volume-limited Main galaxy samples of the Sloan Digital
Sky Survey Data Release 10 (SDSS DR10), I investigate the
environmental dependence of galaxy age, and get the same conclusions
in two volume-limited Main galaxy samples: old galaxies exist
preferentially in the densest regions of the universe, while young
galaxies are located preferentially in low density regions. Such an
age-density relation is likely a combination of a strong age-stellar
mass relation and the stellar mass-density relation.
\end{abstract}


\section{Introduction}\label{sec:intro}
Environmental dependence of galaxy parameters has been an important
subject in the field of galaxy studies. Many galaxy parameters, such
as galaxy luminosity, colour, morphological type, stellar mass and
star formation rate (SFR), exhibit a strong correlation with galaxy
environments (e.g., Postman \& Geller 1984; Dressler et al. 1997;
Hashimoto et al. 1998; Brown et al. 2000; Fasano et al. 2000;
Norberg et al. 2001; Zehavi et al. 2002; Blanton et al. 2003, 2005;
G\(\acute{o}\)mez et al. 2003; Treu et al. 2003; Hogg et al. 2004;
Kauffmann et al. 2004; Li et al. 2006; Zandivarez et al. 2006; Patel
et al. 2009; Deng et al. 2007, 2008a-b, 2009, 2011a, 2012a-b). For
example, Blanton et al. (2003) demonstrated that there is a strong
correlation between luminosity and local density: the most luminous
galaxies tend to reside in the densest regions of the Universe.
Blanton et al. (2005) argued that galaxy colour is the galaxy
property that is most predictive of the local environment. Patel et
al. (2009) reported that the SFR and the specific star formation
rate (SSFR, the star formation rate per unit stellar mass) at z
\(\simeq\) 0.8 show a strong decrease with increasing local density,
similar to the relation at z \(\simeq\) 0. Some studies focused on
the environmental dependence of galaxy age, and concluded that
galaxies in low-density environments are generally younger than
galaxies in high-density environments (e.g. Bernardi et al. 1998;
Trager et al. 2000; Kuntschner et al. 2002; Terlevich \& Forbes
2002; Proctor et al. 2004; Mendes de Oliveira et al. 2005; Thomas et
al. 2005; Gallazzi et al. 2006;
S\(\acute{a}\)nchez-Bl\(\acute{a}\)zquez et al. 2006;
Si\(\acute{l}\)chenko 2006; Reed et al. 2007; Rakos et al. 2007;
Wegner \& Grogin 2008; Smith et al. 2012). For example, Proctor et
al. (2004) and Mendes de Oliveira et al. (2005) reported that the
member galaxies of compact groups are generally older than field
galaxies. Thomas et al. (2005) found that massive early-type
galaxies in low-density regions appear on average \(\simeq\)2 Gyr
younger than their counterparts in high-density regions. Such a
conclusion is in good agreement with the current hierarchical
assembly paradigm, which predicts a younger age of galaxies in lower
density environments (e.g., Lanzoni et al. 2005; De Lucia et al.
2006). However, there also has been the dissenting standpoint.
Kochanek et al. (2000) claimed that the stellar population age of
galaxies does not depend on the environments.

In the SDSS, galaxy spectroscopic targets were selected by two
algorithms. The Main galaxy sample (Strauss et al. 2002) contains
galaxies brighter than \(r_{petro}\) =17.77(r-band apparent
Petrosian magnitude), which is the largest and the most valuable
galaxy sample in the local Universe. The primary goal of this study
is to investigate the environmental dependence of galaxy age in this
sample. The outline of this paper is as follows. Section 2 describes
the data used. In section 3, I presents statistical results for the
environmental dependence of galaxy age in the Main galaxy sample. My
main results and conclusions are summarized in section 4.

 In calculating the
co-moving distance,  I used a cosmological model with a matter
density \(\Omega _0=0.3\), cosmological constant \(\Omega
_\Lambda=0.7\), Hubble's constant
\(H_0\)=100h km s\(^{-1}\) Mpc\(^{-1}\) with h=0.7.

\section{Data}\label{sec:data}
The tenth data release (DR10) (Ahn et al. 2014) of the
SDSS-III is already available. In this work, the data of the Main
galaxy sample was downloaded from the Catalog Archive Server of SDSS
Data Release 10 (Ahn et al. 2014) by the SDSS SQL Search
(http://www.sdss3.org/dr10/). In the SDSS, the target flags can
be used to select out objects that were targeted for some particular
reason.  The Main galaxy targets have one of the LEGACY\_{}TARGET1
bits "GALAXY", "GALAXY\_{}BIG" and "GALAXY\_{}BRIGHT\_{}CORE" set
(bits 6, 7 and 8). This corresponds to the requirement:
LEGACY\_{}TARGET1 \& (64 \(|\) 128 \(|\) 256) $>$ 0. I extract
633172 Main galaxies with the redshift \(0.02\leq z \leq 0.2\). The
data set of age and stellar mass measurements is from the
StellarMassStarFormingPort table obtained with the star-forming
template and the Kroupa IMF (Maraston et al. 2013). I
consider the mass lost via stellar evolution and use best-fit age
of galaxy [in Gyr] and best-fit stellar mass [in log \(M_{sun}\)].

Following Deng (2010), I construct a luminous volume-limited Main
galaxy sample which contains 129515 galaxies at \(0.05\leq z
\leq0.102\) with \(-22.50\leq M_r \leq-20.50\) and a faint
volume-limited sample which contains 34573 galaxies at \(0.02\leq z
\leq0.0436\) with \(-20.50\leq M_r \leq-18.50\). These two
volume-limited Main galaxy samples are located in different redshift
and luminosity ranges. The absolute magnitude \(M_r\) is calculated
from the r-band apparent Petrosian magnitude, using a polynomial fit
formula (Park et al. 2005) for the mean K-correction within 0 $<$ z
$<$ 0.3:
 \[
       K(z) = 2.3537(z-0.1)^2+1.04423(z-0.1)-2.5log(1+0.1).
     \]
     Deng (2010) argued that
when studying the environmental dependence of galaxy properties, one
needs to see the difference between the above-mentioned two galaxy
samples.

\section{Environmental dependence of galaxy age in the Main galaxy sample}\label{sec:results}
The local three-dimensional galaxy density (Galaxies Mpc\(^{-3}\) )
 is defined as the number of galaxies (N=5)
within the three-dimensional distance to the 5th nearest galaxy to
the volume of the sphere with the radius of this distance.  In
previous works (e.g., Deng et al. 2008a, 2009; Deng 2010), such a
density estimator was often applied. In this work, it is still used
to characterize local galaxy environment.  Like Deng et al. (2008a)
did, for each sample, I arrange galaxies in a density order from the
smallest to the largest, select approximately 5\% of the galaxies,
construct two subsamples at both extremes of density according to
the density, and compare distribution of age in the lowest density
regime with that in the densest regime.

Fig. 1 shows age distribution at both extremes of density for the
faint (left panel) and luminous (right panel) volume-limited Main
galaxy samples. As shown by this figure, in these two volume-limited
Main galaxy samples, ages of galaxies strongly depend on local
environments: old galaxies exist preferentially in the densest
regions of the Universe, while young galaxies are located
preferentially in low density regions. I further perform the
Kolmogorov-Smirnov (KS) test. The probability of the two
distributions in Fig. 1 coming from the same parent distribution is
nearly 0, which shows that two independent distributions are significantly
different in this figure. So, this statistical conclusion is robust.

\begin{figure}[t]
\centerline{\includegraphics[width=16cm]{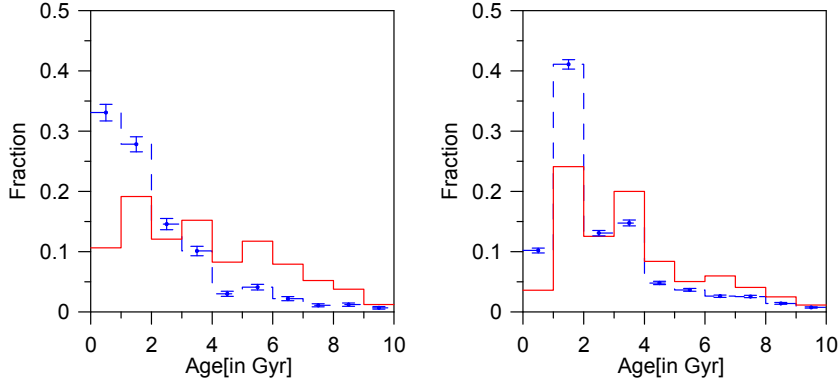}}
\vspace{-16.0cm}
\caption{Age distribution at both extremes of density for the faint
(left panel) and luminous (right panel) volume-limited Main galaxy
samples: red solid line for the subsample at high density, blue
dashed line for the subsample at low density. The error bars of blue
lines are 1-\(\sigma\)  Poissonian errors. Error-bars of red lines
are omitted for clarity.}
\end{figure}

There often are tight correlations between galaxy properties (e.g.,
Bower et al. 1992; {Kennicutt 1992;} Strateva et al. 2001; Blanton
et al. 2003; Hopkins et al. 2003; Baldry et al. 2004; Balogh et al.
2004; Christlein et al. 2004; Kelm et al. 2005; Deng et al. 2008c,
2010; Bamford et al. 2009; Gr\"{u}tzbauch et al. 2011a-b). For
example, Kennicutt (1992) and Bamford et al. (2009) demonstrated
that galaxy morphology is strongly correlated with the SFR and
stellar mass. Deng et al. (2010) reported the correlation between
star formation activities and the concentration index: passive
galaxies are more luminous, redder, highly concentrated and
preferentially ''early-type''. Gr\"{u}tzbauch et al. (2011a) found
that galaxy colour correlates strongly with stellar mass at 0.4 $<$
z $<$ 1. In this condition, the strong environmental dependence of a
galaxy property is likely due to the environmental dependence of
other galaxy properties and tight correlations between galaxy
properties. Gr\"{u}tzbauch et al. (2011b) argued that stellar mass
is the most important factor in determining the colours of galaxies.
Blanton et al. (2005) and Deng and Zou (2009) demonstrated that
galaxy colour is the galaxy property very predictive of local
environments. Thus, the strong environmental dependence of galaxy
age is likely due to the environmental dependence of stellar mass
and tight correlation between stellar mass and age.

Some studies showed that there is a strong correlation between
stellar mass and environment (e.g., Kauffmann et al. 2004; Li et al.
2006; Deng et al. 2011a). Fig. 2 shows stellar mass distribution at
both extremes of density for the faint (left panel) and luminous
(right panel) volume-limited Main galaxy samples. As shown by Fig. 2,
in these two volume-limited Main galaxy samples, high mass galaxies
exist preferentially in the densest regions of the Universe, while
low mass galaxies are located preferentially in low density regions.
The Kolmogorov-Smirnov (KS) test probability in this figure is also
0, which shows that stellar mass of galaxies indeed strongly depends
on environments.

\begin{figure}[t]
\vspace{-0.5cm}
\centerline{\includegraphics[width=16cm]{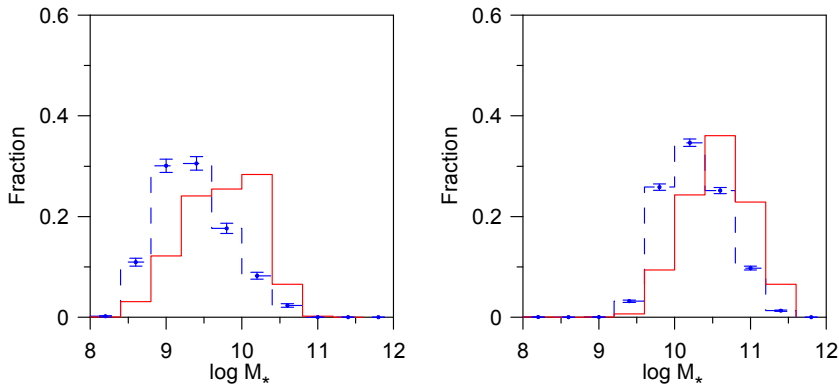}}
\vspace{-16.0cm}
\caption{Stellar mass distribution at both extremes of density for
the faint (left panel) and luminous (right panel) volume-limited
Main galaxy samples: red solid line for the subsample at high
density, blue dashed line for the subsample at low density. The
error bars of blue lines are 1-\(\sigma\)  Poissonian errors.
Error-bars of red lines are omitted for clarity.}
\end{figure}

I examine average stellar mass as a function of age for the faint
(left panel) and luminous  (right panel) volume-limited Main galaxy
samples. Fig. 3 shows that in these two volume-limited Main galaxy
samples, average stellar mass of galaxies increases substantially
with increasing age. This tight age-stellar mass relation and the
stellar mass-density relation likely leads to the strong age-density
relation.

\begin{figure}
\hspace{-1.0cm}
\centerline{\includegraphics[width=16cm]{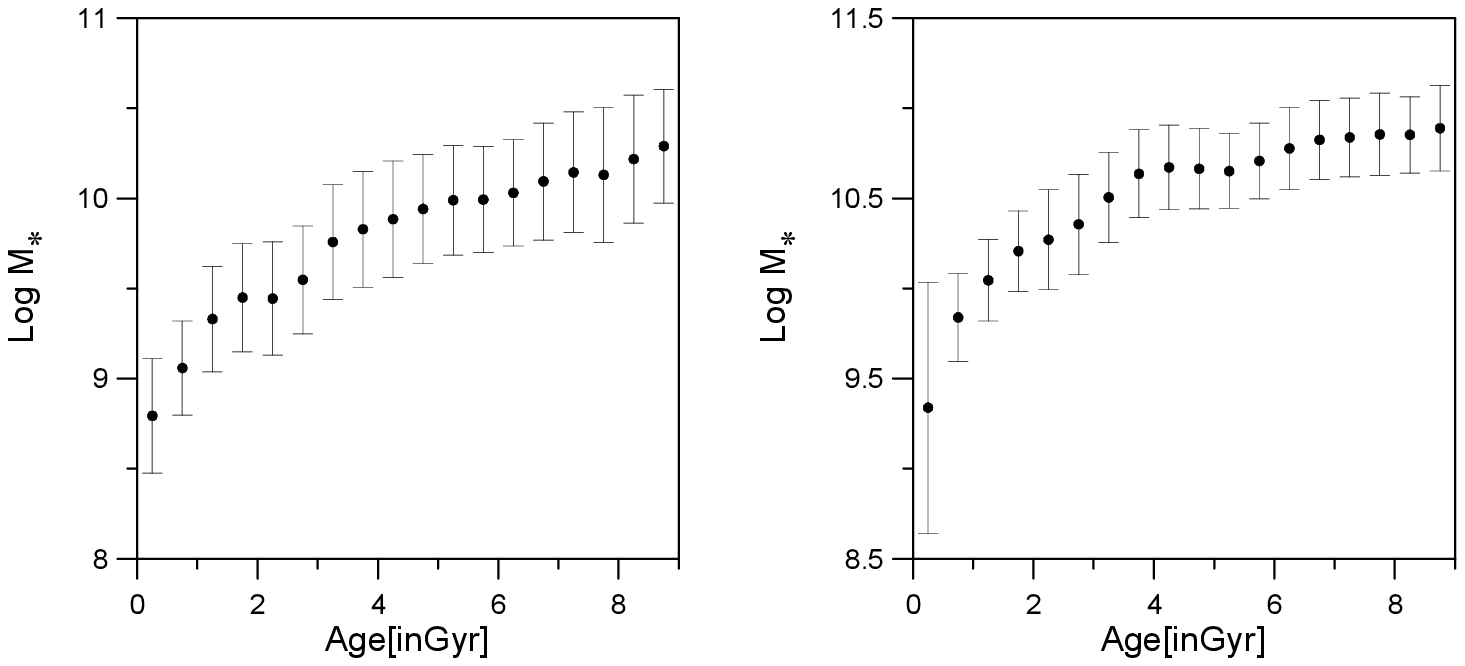}}
\vspace{-16.0cm}
\caption{Average stellar mass as a function of age for the faint
(left panel) and luminous (right panel) volume-limited Main galaxy
samples. Error bars represent standard deviation in each redshift
bin.}
\end{figure}

Norberg et al. (2001) and Deng et al. (2009) demonstrated that the
correlation between the galaxy luminosity and environment is fairly
different between galaxies above and below the value of \(M_r^*\)
found for the overall Schechter fit to the galaxy luminosity
function. Deng et al. (2009) noted that g-r color, concentration
index ci and galaxy morphologies strongly depend on local
environments for all galaxies with different luminosities, and
concluded that \(M_r^*\)  is a characteristic parameter only for the
environmental dependence of galaxy luminosity. Deng et al. (2012c)
even found that \(M_r^*\)  is not an important characteristic
parameter for the environmental dependence of the u-band luminosity.
The u-band luminosity of galaxies still strongly depend on local
environments in the faint volume-limited sample, like the one in the
luminous volume-limited sample does. When investigating the
environmental dependence of other galaxy properties, some works also
demonstrated that there is no significant statistical difference
between galaxies above and below the value of \(M_r^*\) (e.g., Deng
2010; Deng et al. 2011b-c; Deng et al. 2013). In this work, as shown
by Figs. 1-3, the same statistical conclusions can be arrived at
from two volume-limited Main galaxy samples above and below the
value of \(M_r^*\).

\section{Summary}
From the Main galaxy data of SDSS DR10, I construct two
volume-limited samples with the luminosity \(-20.50\leq M_r
\leq-18.50\) and \(-22.50 \leq M_r \leq -20.50\) respectively, and
explore the environmental dependence of galaxy age in these two
volume-limited Main galaxy samples. I apply the three-dimensional
density estimator within the distance to the 5th nearest neighbor,
proceed with the same approach as used by Deng et al. (2008a), and
compares distribution of age in the lowest density regime with that
in the densest regime. Statistical analyses in two volume-limited
Main galaxy samples can reach the same conclusions: old galaxies
exist preferentially in the densest regions of the universe, while
young galaxies are located preferentially in low density regions.
Further investigation suggests that such an age-density relation is
likely a combination of a strong age-stellar mass relation and the
stellar mass-density relation.

\section*{Acknowledgements}

I thank the anonymous referee for providing many useful comments and
suggestions. This study was supported by the National Natural
Science Foundation of China (NSFC, Grant 11263005).

Funding for SDSS-III has been provided by the Alfred P. Sloan
Foundation, the Participating Institutions, the National Science
Foundation, and the U.S. Department of Energy.

The SDSS-III web site is http://www.sdss3.org/. SDSS-III is managed
by the Astrophysical Research Consortium for the Participating
Institutions of the SDSS-III Collaboration including the University
of Arizona, the Brazilian Participation Group, Brookhaven National
Laboratory, University of Cambridge, University of Florida, the
French Participation Group, the German Participation Group, the
Instituto de Astrofisica de Canarias, the Michigan State/Notre
Dame/JINA Participation Group, Johns Hopkins University, Lawrence
Berkeley National Laboratory, Max Planck Institute for Astrophysics,
New Mexico State University, New York University, Ohio State
University, Pennsylvania State University, University of Portsmouth,
Princeton University, the Spanish Participation Group, University of
Tokyo, University of Utah, Vanderbilt University, University of
Virginia, University of Washington, and Yale University.


\begin{thebibliography}{}

\bibitem{Ahn}{Ahn C. P., Alexandroff R., Allende Prieto C., et al., 2014, ApJS, 211, 17}
\bibitem{Baldry} Baldry I. K., Glazebrook K., Brinkmann J., Ivezi\'{c} \u{z}., Lupton R. H., Nichol R. C., Szalay A. S., 2004, ApJ, 600, 681
\bibitem{Balogh} Balogh M. L., Baldry I. K., Nichol R., Miller C., Bower R., Glazebrook K., 2004, ApJ, 615, L101
\bibitem{Bamford}{Bamford S. P., Nichol R. C., Baldry I. K., et al., 2009, MNRAS, 393, 1324}
\bibitem{Bamford}Bernardi M., Renzini A., da Costa L., 1998, ApJ, 508, L143
\bibitem{Blanton}Blanton M. R., Hogg D. W., Bahcall N. A., et al., 2003, ApJ, 594,186
\bibitem{Blanton}Blanton M. R., Eisenstein D., Hogg D. W., Schlegel D. J., Brinkmann J., 2005, ApJ, 629,143
\bibitem{Bower}Bower R. G., Lucey J. R., Ellis R. S., 1992, MNRAS, 254, 601
\bibitem{Brown}Brown M. J. I., Webster R. L., Boyle B. J., 2000, MNRAS, 317, 782
\bibitem{Christlein}Christlein D., McIntosh D. H., Zabludoff A. I., 2004, ApJ, 611, 795
\bibitem{De Lucia}De Lucia G., Springel V., White S. D. M., Croton D., Kauffmann G.,  2006, MNRAS, 366, 499
\bibitem{Deng}Deng X. F., He J. Z., Jiang P., 2007, ApJ, 671, L101
\bibitem{Deng}Deng X. F., He J. Z., Song J., Wu P., Liao Q. H., 2008a, PASP, 120, 487
\bibitem{Deng}Deng X. F., He J. Z., Wu P., 2008b, A\&A, 484, 355
\bibitem{Deng}Deng X. F., He J. Z., Luo C. H., Wu P., Xin Y., 2008c, Acta physica Polonica B,39,965
\bibitem{Deng}Deng X. F., He J. Z., Wen X. Q., 2009, MNRAS, 395, L90
\bibitem{Deng}Deng X. F., Zou S. Y., 2009, APh, 32, 129
\bibitem{Deng}Deng X. F., 2010, ApJ, 721, 809
\bibitem{Deng}{Deng X. F., Yang B., He J. Z., Tang X. X., 2010, ApJ, 708, 101}
\bibitem{Deng}Deng X. F., Chen Y. Q., Jiang P., 2011a, Chinese Journal of Physics, 49, 1137
\bibitem{Deng}Deng X. F., Xin Y., Luo C. H., Wu P., 2011b, Ap, 54, 355
\bibitem{Deng}Deng X. F., Chen Y. Q., Jiang P., 2011c, MNRAS, 417, 453
\bibitem{Deng}Deng X. F., Wu P., Qian X. X., Luo C. H., 2012a, PASJ, 64, 93
\bibitem{Deng}Deng X. F., Song J., Chen Y. Q., Jiang P., Ding Y. P., 2012b, ApJ, 753, 166
\bibitem{Deng}Deng X. F., Song J., Chen Y. Q., Jiang P., Ding Y. P., 2012c, AstL, 38, 213
\bibitem{Deng}Deng X. F., Luo C. H., Xin Y., Wu P., 2013, RMxAA, 49, 181
\bibitem{Dressler}Dressler A., Oemler A. J., Couch W. J., et al., 1997, ApJ, 490, 577
\bibitem{Fasano}Fasano G., Poggianti B. M., Couch W. J., Bettoni D., Kj{\ae}rgaard P., Moles M., 2000, ApJ, 542, 673
\bibitem{Gallazzi}Gallazzi A., Charlot S., Brinchmann J., White S. D. M., 2006, MNRAS, 370, 1106
\bibitem{Gmez}G\(\acute{o}\)mez P. L., Nichol R. C., Miller C. J., et al., 2003, ApJ, 584, 210
\bibitem{Gmez}Gr\"{u}tzbauch  R., Conselice C. J., Varela J., Bundy K., Cooper M. C., Skibba R., Willmer C. N. A., 2011a, MNRAS, 411, 929
\bibitem{Gmez}Gr\"{u}tzbauch R., Chuter R. W., Conselice C. J., Bauer A. E., Bluck A. F. L., Buitrago F., Mortlock A., 2011b, MNRAS, 412, 2361
\bibitem{Hashimoto}Hashimoto Y., Oemler A. J., Lin J. H., Tucker D. L., 1998, ApJ, 499, 589
\bibitem{Hogg}Hogg D. W., Blanton M. R., Brinchmann J., et al., 2004, ApJ, 601, L29
\bibitem{Hopkins}Hopkins A. M., Miller C. J., Nichol R. C., et al., 2003, ApJ, 599, 971
\bibitem{Kauffmann}Kauffmann G., White S. D. M., Heckman T. M.,  M\'{e}nard B., Brinchmann J., Charlot S., Tremonti C., Brinkmann J., 2004, MNRAS, 353, 713
\bibitem{Kelm}Kelm B., Focardi P., Sorrentino G., 2005, A\&A, 442, 117
\bibitem{Kelm}Kennicutt R. C. Jr., 1992, ApJ, 388, 310
\bibitem{Kochanek}Kochanek C. S., Falco E. E., Impey C. D., et al., 2000, ApJ, 543, 131
\bibitem{Kochanek}Kuntschner H., Smith R. J., Colless M., Davies R. L., Kaldare R., Vazdekis A., 2002, MNRAS, 337, 172
\bibitem{Kochanek}Lanzoni B., Guiderdoni B., Mamon G. A., Devriendt J., Hatton S., 2005, MNRAS, 361, 369
\bibitem{Kochanek}Li C., Kauffmann G., Jing Y. P., White S. D. M., B\"{o}rner G., Cheng F. Z., 2006, MNRAS, 368, 21
\bibitem{Kochanek}{Maraston C., Pforr J., Henriques B. M., et al., 2013, MNRAS, 435, 2764}
\bibitem{Kochanek}Mendes de Oliveira C., Coelho P., Gonz\(\acute{a}\)lez J. J., Barbuy B., 2005, ApJ, 130, 55
\bibitem{Kochanek}Norberg P., Baugh C. M., Hawkins E., et al., 2001, MNRAS, 328, 64
\bibitem{Park}Park C., Choi Y. Y., Vogeley M. S., et al., 2005, ApJ, 633,11
\bibitem{Park}Patel S. G., Holden B. P., Kelson D. D.,  Illingworth G. D., Franx M., 2009, ApJ, 705, L67
\bibitem{Kochanek}Postman M., Geller M. J., 1984, ApJ, 281, 95
\bibitem{Kochanek}Proctor R. N., Forbes D. A., Hau G. K. T.,  Beasley M. A., De Silva G. M., Contreras R., Terlevich A. I., 2004, MNRAS, 349, 1381
\bibitem{Kochanek}Rakos K., Schombert J., Odell A., 2007, ApJ, 658, 929
\bibitem{Kochanek}Reed D. S., Governato F., Quinn T., Stadel J., Lake G., 2007, MNRAS, 378, 777
\bibitem{Schawinski}S\(\acute{a}\)nchez-Bl\(\acute{a}\)zquez P., Gorgas J., Cardiel N., Gonz\'{a}lez J. J., 2006, A\&A, 457, 809
\bibitem{Shimasaku}Si\(\acute{l}\)chenko O. K., 2006, ApJ, 641, 229
\bibitem{Sorrentino}Smith R. J., Lucey J. R., Price J., Hudson M. J., Phillipps S., 2012, MNRAS, 419, 3167
\bibitem{Strateva}Strateva I., Ivezic Z., Knapp G. R., et al., 2001, AJ, 122, 1861
\bibitem{Strauss}Strauss M. A., Weinberg D. H., Lupton R. H., et al., 2002, AJ, 124, 1810
\bibitem{Yamauchi}Terlevich A., Forbes D., 2002, MNRAS, 330, 547
\bibitem{Yamauchi}Thomas D., Maraston C., Bender R., Mendes de Oliveira C., 2005, ApJ, 621, 673
\bibitem{Yamauchi}Trager S. C., Faber S. M., Worthey G., Gonz\'{a}lez J. J., 2000, AJ, 120, 165
\bibitem{Yamauchi}Treu T., Ellis R. S., Kneib J., Dressler A., Smail I., Czoske O., Oemler A., Natarajan P., 2003, ApJ, 591, 53
\bibitem{Yamauchi}Wegner G., Grogin N. A., 2008, AJ, 136, 1
\bibitem{Yamauchi}Zandivarez A., Mart\(\acute{i}\)nez H. J., Merch\(\acute{a}\)n M. E., 2006, ApJ, 650, 137
\bibitem{Yamauchi}Zehavi I., Blanton M. R., Frieman J. A., et al., 2002, ApJ, 571, 172

\end{thebibliography}
\end{document}